\documentclass[aps,prl,twocolumn,groupedaddress,showpacs]{revtex4}

\usepackage{graphicx}
\usepackage{dcolumn}
\usepackage{bm}

\begin{document}


\title{Angle-dependent ultrasonic transmission through plates with subwavelength hole arrays}

\author{H\'ector Estrada}
\author{Pilar Candelas}
\author{Antonio Uris}
\author{Francisco Belmar}
\author{Francisco Meseguer}
\email[To whom correspondence should be addressed. ]{fmese@fis.upv.es}

\affiliation{Centro de Tecnolog\'ias F\'isicas, Unidad Asociada ICMM- CSIC/UPV, Universidad Polit\'ecnica de Valencia, Av. de los Naranjos s/n. 46022 Valencia, Spain}

\author{Francisco J. \surname{Garc\'ia} de Abajo}
\affiliation{Instituto de \'Optica - CSIC Serrano 121, 28006 Madrid, Spain}
\date{\today}

\begin{abstract}
We study sound transmission in perforated plates as a function of incident angle and conclude that it holds distinctive properties that make it unique and essentially different from optical transmission through perforated metallic plates. More precisely, we conclude the following: (a) similar to its optical counterpart, acoustic transmission minima respond to Wood anomalies in which the periodicity plays a central role; (b) in contrast to both the optical case and the acoustical case with slits, homogeneous-plate modes (Lamb and Scholte-Stoneley modes) are strongly coupled to lattice and Fabry-P\'erot resonances. This gives rise to unique transmission behavior, thus opening new perspectives for exotic wave phenomena.
\end{abstract}

\maketitle

The similarities and discrepancies between mechanical waves such as sound and electromagnetic waves have puzzled scientists for a long time \cite{brillouin1953}. This is much more than a semantic question. It deals with the essential nature of  matter since sound is basically a classical physics phenomenon while light is deeply rooted into the quantum mechanical description of nature. Light and sound have been confronted in numerous experiments and theories, ranging from band gap effects in photonic \cite{krauss1996} and phononic \cite{martinezsala1995} crystals to negative refraction effects \cite{ruan2006,feng2005} and invisibility \cite{Cai2007,cummer2007}. The discovery of extraordinary optical transmission effects in metallic membranes perforated by subwavelength apertures has again raised the question of the similarity between sound and light in the context of wave transmission. Very recently several groups have reported on the transmission properties of sound through plates with  1D \cite{christensen2007,lu2007} and 2D \cite{hou2007,estrada2008} apertures. Some groups \cite{christensen2007,lu2007} claim that extraordinary transmission of sound is similar to the optical counterpart in corrugated metal films. However, Hou et al. \cite{hou2007}, have shown that although same similarities appear, intrinsic differences separate light and sound. Holes in membranes, can not sustain optical modes in the subwavelength region. This is not the case for sound. Moreover we have recently shown \cite{estrada2008} that sound has specific properties unforeseen from the perspective of optical transmission in metallic films. Firstly, extraordinary transmission effects are not directly related to hole periodicity since it also appears for perforated plates with a random distribution of holes. More importantly, periodically perforated plates are capable of shielding sound, in the region of the Wood anomaly,  much better than what is predicted by the well known mass law \cite{estrada2008}. This behavior is unique and different from optics, and it can lead to killing applications in sound proofing engineering. However, a detailed study is still missing regarding the role of the wavevector parallel to the plate in sound transmission for oblique incidence in both periodically and random distributed holes.
In this Letter, we investigate the angular dependence of sound transmission through plates perforated by subwavelength apertures. Our results reinforce our previous findings and demonstrate that sound in perforated plates behaves like light in high-refractive-index materials. We find that elastic Lamb waves are strongly coupled to lattice modes induced by the periodical distribution of holes. Fabry-P\'{e}rot modes are also intervening in the transmission, giving rise to a rich behavior intrinsic to the mechanical character of acoustic waves.
\begin{figure*}
 \includegraphics{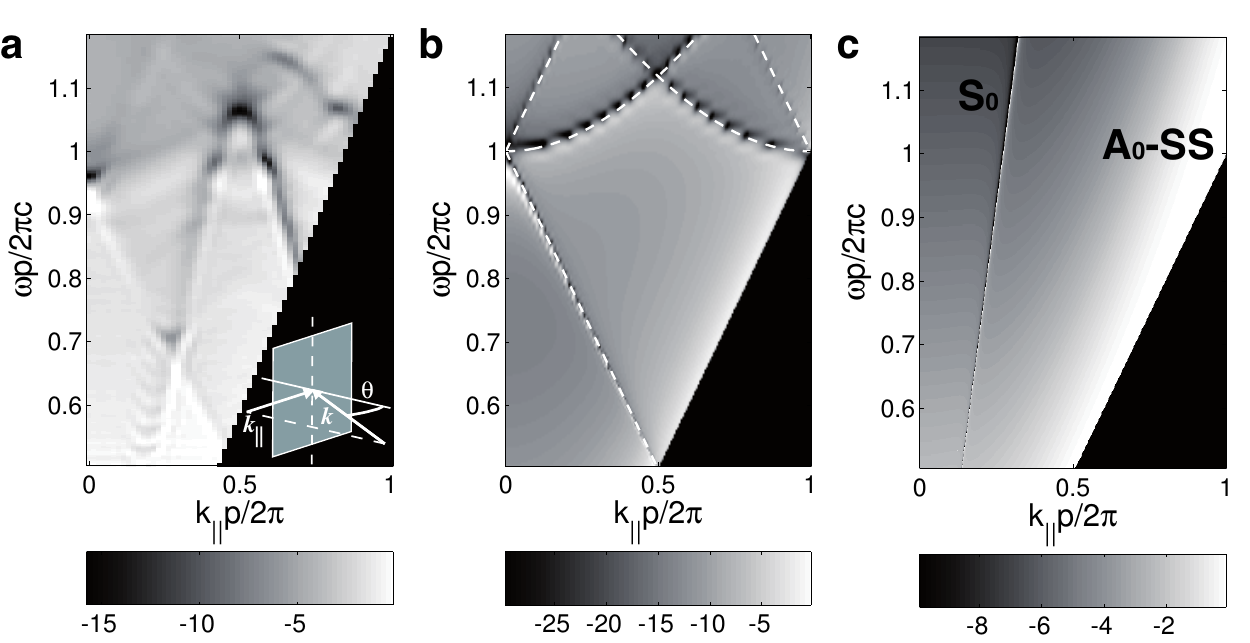}%
 \caption{\label{t1} Transmission (gray scale in dB) as a function of parallel momentum $k_{||}$ and frequency $\omega$ for a perforated plate with $d=3$ mm, $p=5$ mm, and $t=1$ mm from (a)  measurements and (b) hard-solid calculations. Dotted white lines in (b) represent the Wood anomalies given by the Eq. (\ref{wood}). (c) Transmission calculations for a plate without holes and the same thickness.}
 \end{figure*}
As already reported \cite{estrada2008}, our experimental setup is based on the well known ultrasonic immersion transmission technique. The angle of incidence $\theta$ has been varied by rotating the sample plate from 0$^\circ$ to 60$^\circ$ in steps of 1$^{\circ}$. The plates are made of Al ($\rho = 2.7$ g/cm$^3$,  $c_{l} = 6500$ m/s, and $c_{t} = 3130$ m/s) \cite{ultranondes}, drilled mechanically, and immersed in water. The plates measured in this work have a hole diameter $d=3$ mm and are distributed periodically in a squared array with a periodicity $p=5$ mm. The measured transmission spectra are collected as a function of the wave vector parallel to the plate $\mathbf{k}_{||}$ in the $\overline{\Gamma X}$ direction of the reciprocal lattice, and the frequency $\omega$. The former is related to both frequency and angle by $|\mathbf{k}_{||}|=k\sin\theta$ with $k = \omega/c$, where $c$ is the sound speed in water and $\theta$ is the angle of incidence as shown in the inset of Fig. \ref{t1}(a).

The model used to calculate the transmission of the perforated plates \cite{estrada2008} is based on an expansion of the acoustic field in terms of modes of a cylindrical cavity inside the hole and plane waves outside the hole, where the expansion coefficients are determined by imposing continuity of the pressure and velocity fields on both planes limiting the plate. No field is considered inside the plate, which is equivalent to the perfect conductor idealization for electromagnetic waves. On the other hand, for the plate without holes we used a simple fluid-solid-fluid model \cite{ultranondes}.

The Fig. \ref{t1}(a) shows the transmission intensity map in dB as a function of the normalized $k_{||}$ and the reduced frequency ($\omega p/2\pi c$) of a perforated plate with $d=3$ mm, $p=5$ mm, and thickness $t=1$ mm. Figure \ref{t1}(b) shows the calculated transmission intensity  for the same sample as in Fig. \ref{t1}(a). For $k_{||}$ near zero (normal incidence) the Wood anomaly can be clearly seen both in the experiments and the calculations at $\omega p/2\pi c \approx 1$. However, the measured anomaly [Fig. \ref{t1}(a)], it is slightly shifted to lower frequencies with respect to the numerical one, which we attribute to the finite size of the plate and assumed infinite impedance mismatch of the water/plate interface. For this case, the acoustic transmission exhibits sharp features in both theory and experiment, although the latter is embedded in a lighter grey background representing finite transmission that is facilitated by penetration of the sound inside the solid. Figure \ref{t1}(c) shows the calculation of the plate modes when no holes are drilled. Here, the transmission is dominated by the presence of the S$_0$ and A$_0$ Lamb modes \cite{ultranondes}, and the latter is mixed with the Scholte-Stoneley mode near grazing incidence. The measurement in the drilled plate [see Fig. \ref{t1}(a)] resembles the hard solid theory, but a new mode consistent with S$_0$ appears and strongly interacts with the Wood anomaly at $k_{||}p/2\pi =0.3 ,\ 0.5$.
In our case, the condition for the Wood anomaly can be written as
\begin{equation}
    |\mathbf{k}| = |\mathbf{k}_{||} + \mathbf{g}| = \sqrt{(k_{||} + 2\pi n/p)^2 + (2\pi m/p)^2}\,,\label{wood}
\end{equation}

where $\mathbf{k}$ is the incident wavevector and $\mathbf{g}$ corresponds to the lattice momentum for the square array with integers $n$ and $m$. This expression is related to the singularities of the structure-factor of the lattice \cite{abajo2007}, which originate in accumulative in-phase scattering in the hole array and gives rise to complementary phenomena for electromagnetic waves in hole and particle-arrays via Babinet's principle \cite{abajo2005E}. The different values for $n$ and $m$ produce different white dashed lines in Fig. \ref{t1}(b), which is in excellent agreement with the hard-solid model and the measurements.

\begin{figure*}[ht]
 \includegraphics{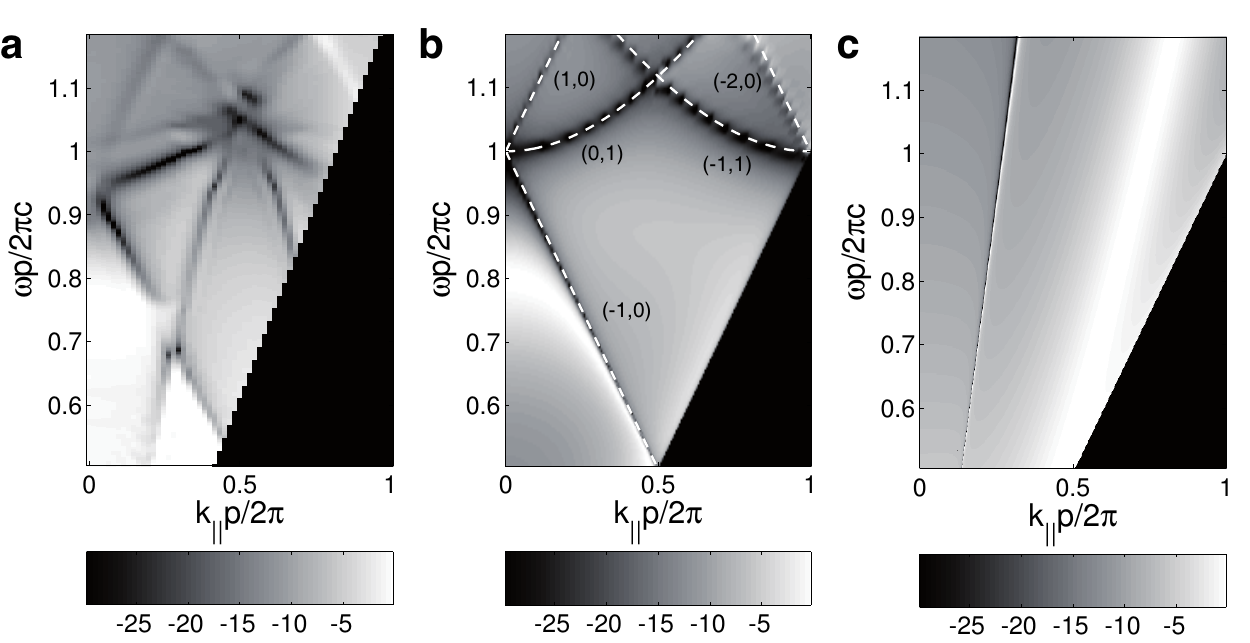}%
 \caption{\label{t2} Transmission (gray scale in dB) as a function of parallel momentum $k_{||}$ and frequency $\omega$ for a perforated plate with $d=3$ mm, $p=5$ mm, and $t=2$ mm from (a)  measurements and (b) hard-solid calculations. Dotted white lines in (b) represent the Wood anomaly given by the Eq. (\ref{wood}) and are labeled as $(n,m)$. (c) Transmission calculations for a plate without holes and the same thickness.}
 \end{figure*}
Even better agreement can be observed in Fig. \ref{t2}(a), (b), where the plate thickness has been increased up to 2 mm. We have made both experiments and calculation for reference undrilled plates. The features in this case correspond to plate modes as discussed above for the former sample (see Fig. \ref{t1}). Here again Lamb modes [shown in Fig. \ref{t2}(c) for a homogeneous plate], which are not taken into account in the calculation shown in figure \ref{t2}(b), interact with Wood anomaly modes. Also in this case the sound penetration into the solid excites a mode that cross the $(-1,0)$ Wood anomaly, meets the $(0,1)$, $(-1,1)$ crossing, and falls down with negative group velocity outside the first Brillouin zone. However closer examination of the results shown across all panel in Fig \ref{t2} indicates more complicated interplay between hole-induced modes and homogeneous plate modes. Firstly, the observed Wood anomaly lines are slightly shifted to lower frequency values with respect to the calculated ones. Secondly, the S$_0$ mode [see Fig. \ref{t2}(a)] in the perforated plate is significantly less steep than in the non perforated plate [Fig. \ref{t2}(c)]. This means that the phase velocity of the S$_0$ plate mode is smaller when holes are drilled in the plate.
\begin{figure*}
 \includegraphics{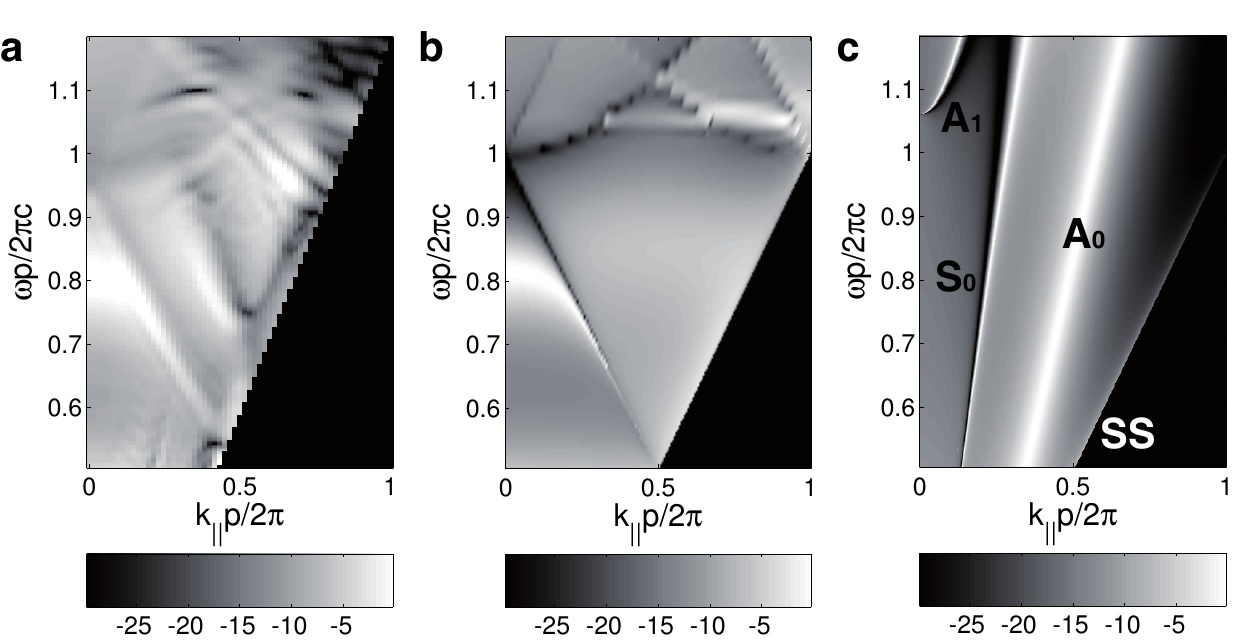}%
 \caption{\label{t5} Transmission (gray scale in dB) as a function of parallel momentum $k_{||}$ and frequency $\omega$ for a perforated plate with $d=3$ mm, $p=5$ mm, and $t=5$ mm from (a) measurements, (b) hard-solid calculations, and (c) analytical calculations \cite{ultranondes} for the non-perforated plate and the same thickness.}
 \end{figure*}

As the thickness increases, the interaction between Lamb and hard-solid modes becomes more involved. In order to understand the experiments, let us first discuss the plate modes of reference samples. For $t=5$ mm, the A$_0$ Lamb mode moves to lower momentum and frequency values, away from the Scholte-Stoneley mode. Moreover, the A$_1$ Lamb mode appears in the range of the measurements. The transmission due to the hole array [see fig \ref{t5}(a)] has two peaks due to enhanced Fabry-P\'erot resonances, with the second resonance showing up in the range of the measurements, flanked by the lower Wood anomaly line, while the first resonance at $\omega p/2\pi c = 0.42$ is out of the observation range. The S$_0$-like mode that emerged for small thickness does not exceed $\omega p/(2\pi c)=0.6$. Instead, many modes with high transmission values can be seen above the Wood anomaly $(-1,0)$. The full transmission due to the holes becomes broader as compared to smaller thicknesses. Similar to the preceding samples, an antimode that is not predicted by the hard-solid model crosses the lower Wood anomaly but in this case the tail with negative group velocity cannot be clearly resolved. A theory taking into account the influence of plate modes will be required to shed light into the complex interplay between Wood anomalies and plate modes, although that lies outside the scope of the present paper.
\begin{figure}
 \includegraphics{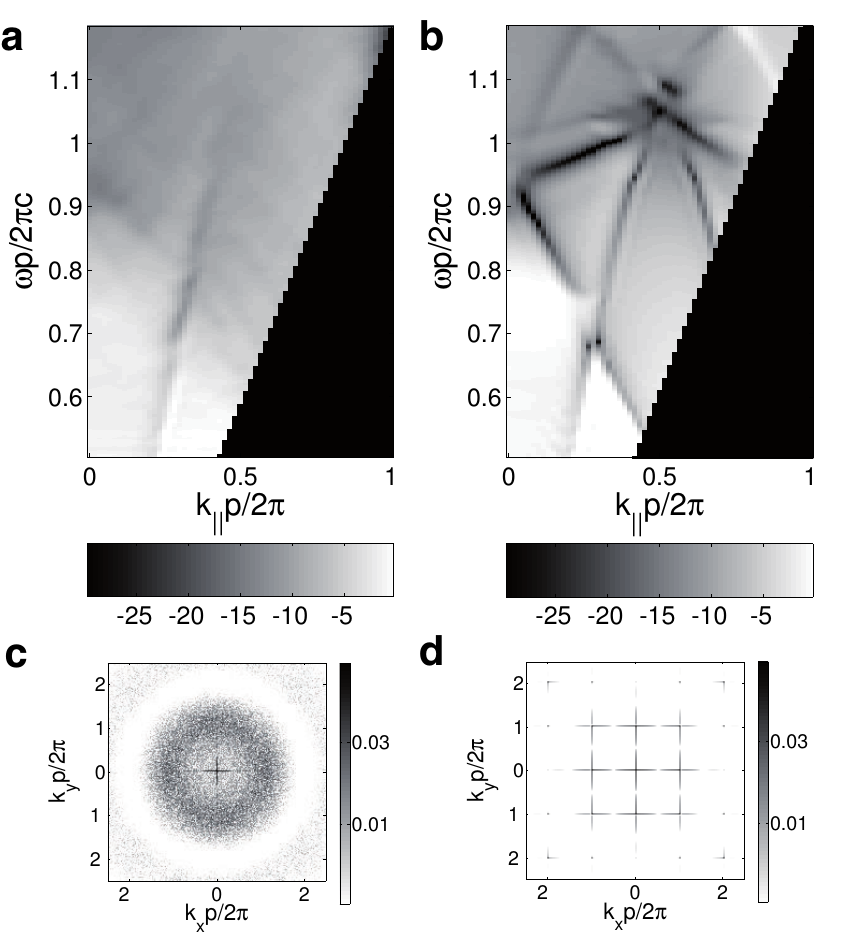}%
 \caption{\label{al} Transmission measurements (gray scale in dB) as a function of parallel momentum $k_{||}$ and frequency $\omega$ for a perforated plate with $d=3$ mm, $t=2$ mm, and $p=5$ mm (a) compared to a plate of the same thickness and randomly distributed holes (b). The filling fractions of the holes is the same in both plates. 2D Fourier transforms (contour plots in log scale) of the (c) random and (d) periodic array are shown as well.}
 \end{figure}

The dramatic influence of hole ordering on the transmission performance of perforated plates is shown in Figs. \ref{al}(a), (b). It corresponds to the experimental transmission results of plates of 2 mm in thickness, perforated randomly [Fig. \ref{al}(a)] and periodically with $p=5$ mm [Fig. \ref{al}(b)], both arrays having the same filling fraction of apertures and the same hole diameter ($d=3$ mm). The very rich interplay between the lattice modes and those of the plate [Fig. \ref{al}(a)] is completely absent when the holes are arranged randomly. The random array exhibits only a small dark zone that slightly resembles the Wood anomaly for small $k_{||}$, accompanied by a plate mode resembling the S$_0$ Lamb mode of the homogeneous plate [see Fig. \ref{al}(a)]. Furthermore, some other Lamb modes of the homogeneous plate are suppressed by the presence of a non-periodic hole array [compare Figs. \ref{t2}(c) with \ref{al}(a)]. The phase-velocity of the S$_0$-like mode is found to be 3148 m/s, which is very close to the nominal value of the bulk transversal wave (3130 m/s) and slower than the S$_0$ mode of non-perforated plate (5486 m/s for low frequency) [see figure \ref{t2}(c)]. Then, this mode is independent of the lattice-period and one can expect that the interaction of these modes with those resulting from the periodicity are the basis of the phenomena observed in Figs. \ref{al}(b) and in Fig. \ref{t5}(a), although the latter involves more Lamb modes in the interaction. For the periodic case, the phononic crystal generated for the Lamb waves \cite{zhangxinya2006} could interact with the Fabry-P\'erot resonances and lattice-driven-modes as well, but further study is required to address this question.

In summary, the dips observed in the acoustical transmission through perforated plates have a similar origin as those in the transmission of light in drilled metallic films. These features are strongly connected to the Wood anomalies, as we demonstrate with measurements and calculations. The finite impedance contrast between the fluid and the plate plays a central role, although a small penetration of the sound into the bulk of our metallic plates produces a background transmission that produces an interesting interplay between plate modes and direct bulk transmission. Intrinsic elastic modes of homogeneous plates (Lamb modes) couple to Wood-anomaly-driven modes and Fabry-P\'erot modes, thus generating a new scenario in the context of wave transmission through subwavelength apertures. Our calculations, based upon the hard-solid approximation, are capable of explaining the role of Wood anomalies and their corresponding lattice resonances in the transmission features. However, our measurements reveal a major role played by Lamb modes, which have been shown to interact with lattice modes, leading to a reduction in the phase velocity of the former as a result of this interplay. We hope that these effects will trigger further experimental work to further explore the possibilities of this interplay, as well as more realistic theoretical descriptions incorporating intrinsic plate modes and Wood anomalies on a single footing, although that is clearly outside the scope of the present paper.

\begin{acknowledgments}
This work has been partially supported by the Spanish CICyT (projects  MAT2006-03097; MAT2007-66050, and Consolider CSD2007-00046), Intramural Project  CSIC Nr.:200560F0071. H. Estrada acknowledges CSIC-JAE scholarship.
\end{acknowledgments}


\end{document}